\documentclass[%
 showkeys,%
 amsmath,amssymb,%
]{revtex4-2}

\usepackage{graphicx}
\usepackage{dcolumn}
\usepackage{bm}
\usepackage{float}
\usepackage[hyperindex,colorlinks,linkcolor=cyan,citecolor=magenta ]{hyperref}
\usepackage[utf8]{inputenc}
\usepackage[T1]{fontenc}
\usepackage{physics}

\begin{document}


\title[Sample title]{Quantization of a Black-Hole Gravity: Geometrodynamics and the Quantum Phase-Space Formalism}


\author{Carla R. Almeida}
  \email{cralmeida00@gmail.com}
 \affiliation{Institute of Physics, University of São Paulo (IFUSP).}
\author{Denis C. Rodrigues}%
 \email{deniscr@gmail.com}
\affiliation{International Ph.D. Program in Astrophysics, Cosmology and Gravitation, Federal University of Espírito Santo (PPGCosmo - UFES).
}%


\date{\today}

\begin{abstract}
Quantum gravity is effective in domains where both quantum effects and gravity are essential, such as in the vicinity of space-time singularities. This paper will investigate the quantization of a black-hole gravity, particularly the region surrounding the singularity at the origin of the coordinate system. Describing the system with a Hamiltonian formalism, we apply the covariant integral quantization method to find the Wheeler-DeWitt equation of the model. We find that the quantized system has a discrete energy spectrum in the region inside the event horizon. Through the Kantowski-Sachs metric, it is possible to correlate the entropic time, which gives the dynamics for this model, to the cosmic time in a non-trivial way. Different configurations for the phase space of a Schwarzschild black hole are obtained in a semi-classical analysis. For lower-energy states, the quantum corrections result in singularity removal and wormhole formation.
\end{abstract}

\maketitle

\section{Introduction}

With the detection of gravitational waves produced by the merge of two black holes and the release of a shadow-image by the Event Horizon Telescope team, it seems safe to say that the near future of black-hole physics is exciting. New observational techniques and technologies provide promising tools to learn more about those objects, and the theory tells us there is still plenty to unveil about them. Perhaps the most curious feature of those objects is the singularity hidden by the event horizon---a hole of infinity density in the fabric of space-time. Strange as it is, this is a proper prediction of general relativity (GR) \cite{Hawking.1970}. The singularity is, however, located in a region forever outside our reach. The interior of a black hole is inaccessible to external observers. Nevertheless, there is still a chance to test theories about that region via measurable phenomena, like gravitational waves and Hawking radiation.

The investigation of black holes using quantum theory is not new. In fact, the thermodynamical theory proposed in the early 1970s \cite{Hawking.1974} requires the consideration of quantum effects near the event horizon. The imbalance of particle/anti-particle production in this region leads to a low but constant rate of produced particles emitted to infinity. This phenomenon became known as Hawking radiation, named after the physicist who first proposed it. In the extreme environment of the neighborhood around the singularity, however, gravity virtually annuls any quantum effects if we consider this same \textit{ad hoc} merge of quantum theory and GR. An alternative would be to quantize gravity itself to investigate if the quantization leads to relevant corrections to the classical theory. Naturally, quantum gravity has a specific scope of action: Planck-scale regions with strong gravitational influence.

This paper investigates precisely this region of a Schwarzschild black hole, using quantum-gravity theory to explore the implications of the Wheller-DeWitt equation. The canonical quantization of spherical symmetric metrics has been extensively researched for a while now \cite{Thiemann.1993,Obregon.1998, Yajnik.1998,Bouhmadi.2020,Casadio.2022}, and the topic has also been explored through the lenses of loop quantum theories \cite{Campiglia.2007,Modesto.2008,Gambini.2013,Bojowald.2020}. Our approach, however, differs from those of previous works. Instead of the canonical, we will consider another quantization method based on the symmetry group of the system's phase space: the covariant integral method.

The canonical method, based on the operators
\begin{equation}
\label{canonical_op.}
    \hat{Q} : f(x) \mapsto x f(x) \quad ; \quad \hat{P}: f(x) \mapsto -i \frac{d}{dx} f(x) \,,
\end{equation}
is more suitable for variables $x \in \mathbb{R}$. However, it becomes inadequate for different coordinate domains \cite{Twareque_Ali.2005} since the quantization depends on the symmetry between position and momentum, and different domains have different symmetry groups for their phase space. The covariant integral method is, thus, a more interesting alternative than the canonical one. In addition, the integral method takes into consideration those symmetries to obtain quantum observable for any classical function or distribution of the position and momentum. It is worth mentioning that the canonical operators \eqref{canonical_op.} can be obtained as a particular case of the covariant integral quantization for variables in the whole real line, in which the symmetry is given by the Weyl-Heisenberg group \cite{Gazeau.2014}.

Gravitational models, including cosmological ones, are often described by strictly positive parameters, such as the radius in the polar coordinate or the scale factor of the Universe in an FLRW metric. The canonical quantization of this type of variable is often clunky, leading to the necessity of extra conditions and terms placed by hand to maintain the desirable features of a quantum theory, for example, the self-adjointness of the kinetic potential operator. To avoid these issues in the quantization of strictly positive variables, $\mathbb{R}_{+}^{*} = \{x \in \mathbb{R} | x > 0\}$, we shall use the covariant integral method based on the affine group,
\begin{eqnarray}
\label{aff_group}
    \Pi_{+} := \{(q,p) | q>0 ; p\in \mathbb{R} \} \quad ; \quad (q,p) (q_{0}, p_{0} ) := \left( qq_{0}, \frac{p_{0}}{q} + p \right) \,.
\end{eqnarray}

The covariant integral method is part of a quantum phase-space (QPS) formalism that has a long tradition in quantum field theory research, including applications to quantum optics, quantum information, and quantum technologies in general \cite{Aldrovandi.1990, Rundle.2021}. The affine quantization method, in particular, has been recently applied to early-Universe models \cite{Bergeron.2014,Bergeron.2015,Frion.2019}, resulting in the replacement of the initial singularity for a smooth bounce in the semi-classical limit. For the case of black holes, affine quantization was used in simplified models and had interesting results, such as singularity removal and displacement of the horizon \cite{Zonetti.2014}.

We will approach the problem from a different angle than \cite{Zonetti.2014}. Our goal is to study the quantum system described by the Wheeler-DeWitt (WdW) equation derived from a classical black-hole model. The Hamiltonian description of general relativity is a constrained system \cite{Gourgoulhon.2012}. In particular, the quantization of the Hamiltonian constraint, $H = 0$, leads to the WdW equation \cite{DeWitt.1967} 
\begin{equation}
    \hat{H} \Psi= 0 \,.
\end{equation}
When compared to the Schr{\"o}dinger equation, which governs the dynamics of the quantum system, we notice that the wave function $\Psi$ in \eqref{WdW} is time-independent. It is known as the problem of time \cite{Isham.1993}: time disappears in quantized gravitational models. It happens mainly because the Hamiltonian constraint is valid in the spatial hypersurface of a 3+1 decomposition,
\begin{equation}
    g_{\mu \nu} dx^{\mu} dx^{\nu} = - N^{2} dt^{2} + \gamma_{ij} (dx^{i} + \beta^{i} dt)(dx^{j} + \beta^{j} dt) \,,
\end{equation}
and thus it is effectively calculated ``for a given time.'' In the ADM formalism, the functions $N$ and $\beta^{i}$ are not dynamical entities. Therefore, we shall consider that another intrinsic parameter is responsible for the dynamics of the quantized system, establishing an entropic dynamic to the model \cite{Caticha.2011}. We will investigate how this realizes in the case of black holes.

For this initial investigation, we will consider $c=\hbar = G =1$.

This paper is organized as follows: Section \ref{AffQuant} presents the tools to apply the affine quantization method in the model. In Section \ref{Hamilt_descrip}, we review the Hamiltonian formalism for a black hole, as laid out in Cavaglià et al. \cite{Cavaglia.1996}. The quantization of the black-hole system via the Wheeler-DeWitt equation is presented in Section \ref{BH_quant}. In Section \ref{Interior_BH}, we restrict our calculations to the interior of the event horizon, where we introduced a brief review of the Kantowski-Sachs metric for this region \cite{Kantowski.1963}. Finally, the semi-classical limit offered by the affine quantization method is presented in Section \ref{QPS portrait}. We conclude our paper with a discussion of our results.

\section{An exposition of the affine quantization method}
\label{AffQuant}

The affine quantization method has a rich mathematical foundation. A more profound revision of this method can be found in \cite{Almeida.2018}. For our purpose, however, we will present only a brief exposition of the quantization procedure and dequantization map. The goal is to pose enough formalism to understand the tools we will use, that is, the final form of the affine operator and the introduction of a semi-classical picture derived from the method.

\subsection{Quantization Map}

Let us consider the Hilbert space $\mathcal{H}_{\gamma}$, defined by
\begin{equation}
\label{Hilbert space}
    \mathcal{H}_{\gamma} := L^{2} \left( \mathbb{R}_{+} \, , \, \frac{dx}{x^{\gamma+1}}\right) \,.
\end{equation}
This is the space of square-integrable functions on the half-plane equipped with the measure $d\mu(x) = dx/x^{\gamma+1}$. It has a continuous basis $|x \rangle$, which is orthogonal and closed, that is,
\begin{equation}
    \langle x | x^{\prime} \rangle = x^{\gamma +1} \delta(x-x^{\prime}) \quad; \quad \int_{0}^{\infty} \frac{dx}{x^{\gamma +1}} | x \rangle \langle x^{\prime} | = I_{\mathcal{H}_{\gamma}} \,,
\end{equation}
where $I_{\mathcal{H}_{\gamma}}$ is the identity of $\mathcal{H}_{\gamma}$. To build the quantum operator, we will use of the so-called \textit{coherent states} $|q,p\rangle$, quantum states representing the physical system \cite{Gazeau.2009}, defined as
\begin{equation}
    \langle x| q,p \rangle = q^{\frac{\gamma}{2}} e^{ipx} \psi \left( \frac{x}{q} \right) \,,
\end{equation}
where
\begin{equation}
\label{psi}
    \psi (x) \in L^{2} \left( \mathbb{R}_{+} \, , \, \frac{dx}{x^{\gamma + 1}}\right) \cap L^{2} \left( \mathbb{R}_{+} \, , \, \frac{dx}{x^{\gamma+2}}\right) = \mathcal{H}_{\gamma} \cap \mathcal{H}_{\gamma + 1} \,.
\end{equation}
The function $\psi$ is called a \textit{fiducial vector} or \textit{wavelets}, and the condition \eqref{psi} is necessary to ensure the smoothness of the quantum operator. In this paper, we will consider $\psi: \mathbb{R}_{+} \rightarrow \mathbb{R}$. Thus, $\overline{\psi} = \psi$.

The affine quantization is a map that takes a classical function $f(q,p)$ and turns it into a unique operator $A_{f}$ such that
\begin{equation}
\label{Af_def}
    A_{f} = \int_{0}^{\infty} \int_{-\infty}^{\infty} \frac{dqdp}{2\pi c_{\gamma}} f(q,p) |q,p\rangle \langle q,p| \,.
\end{equation}
The constant $c_{\gamma} = c_{\gamma}^{(0)}$ is defined as
\begin{equation}
    c_{\gamma}^{(n)} = c_{\gamma}^{(n)}(\psi) = \int_{0}^{\infty} \frac{dx}{x^{\gamma + 2}} |\psi^{(n)}(x)|^{2} \,,
\end{equation}
where $\psi^{(n)}$ is the $n^{th}$ derivative of $\psi$. Therefore, for a generic function $\varphi (x): \mathbb{R}_{+} \rightarrow \mathbb{R}$, the operator $A_{f}$ applied to $\varphi$ in the basis $|x \rangle$ is given by,
\begin{equation}
\label{Af_formulae}
     \langle x| A_{f} | \varphi \rangle = \int_{0}^{\infty} \int_{-\infty}^{\infty} \frac{dqdp}{2\pi c_{\gamma}} f(q,p) \int_{0}^{\infty} \frac{dx^{\prime}}{(x^{\prime})^{\gamma + 1}} q^{\gamma} e^{ip(x-x^{\prime})} \psi \left( \frac{x}{q} \right) \psi \left( \frac{x^{\prime}}{q} \right) \varphi(x^{\prime}) \,.
\end{equation}
$A_f$ is the equivalent quantum observer of the classical function $f$.

Clearly, the covariant integral method, which we are presenting here for the affine group \eqref{aff_group}, is quite different from the canonical one. However, we can better understand its realization with the quantization of classical parameters, such as position, momentum, and kinetic energy.

Using \eqref{Af_formulae}, the quantization of $q^{\beta}$ and $p$ yields,
\begin{eqnarray}
\label{quantiz_q}
\langle x | A_{q^{\beta}} | \varphi \rangle = \frac{c_{\gamma + \beta}}{c_{\gamma}} x^{\beta} \varphi(x) \quad &\Rightarrow& \quad A_{q^{\beta}} =\frac{c_{\gamma + \beta}}{c_{\gamma}} \hat{Q}^{\beta} \,;
\\
\label{quantiz_p}
\langle x | A_{p} | \varphi \rangle = -i\frac{d}{dx} \varphi(x) + i \left(\frac{\gamma+1}{2} \right) \frac{1}{x} \varphi(x) \, &\Rightarrow& \, A_{p} = \hat{P} + i \left(\frac{\gamma+1}{2} \right) \hat{Q}^{-1} \,;
\end{eqnarray}
where $\hat{Q}$ and $\hat{P}$ are, respectively, the canonical position and momentum operators defined in \eqref{canonical_op.}. Writing $A_{q^{\beta}}$ and $A_{p}$ in terms of the canonical operator allows us to visualize how the affine method acts. Constant aside, the affine quantization of the position results in the same canonical operator for this parameter. However, there is a more dramatic change for the momentum quantization with the appearance of a term $\hat{Q}^{-1}$. This happens because we are considering a general Hilbert space $\mathcal{H}_{\gamma}$ for the system. If, for example, $\gamma = -1$, that is, $\mathcal{H}_{-1} = L^{2}(\mathbb{R}_{+}, dx)$, then $A_{p}$ coincides with the canonical momentum operator. With a similar calculation, we can find $A_{p^{2}}$,
\begin{equation}
\label{quantiz_p2}
A_{p^{2}} = \hat{P}^{2} + i \left(\gamma+1\right) \hat{Q}^{-1} \hat{P} + \left( \frac{c_{\gamma-2}^{(1)}}{c_{\gamma}} - \frac{(\gamma+1)(\gamma +2)}{2} \right) \hat{Q}^{-2} \,. 
\end{equation}
We see that, at least for $\gamma =-1$, the affine quantization naturally recovers the self-adjoint character of the kinetic potential.

\subsection{Dequatization Map}
\label{dequantization}

Definition \eqref{Af_def} provides an interesting way to obtain a dequantization map. Given an observer $\hat{\mathcal{O}}(q,p)$, we can recover a classical observer $\check{f}_{\mathcal{O}}(q,p)$ as the expected value of the observer with respect to the coherent states $|p,q \rangle$,
\begin{equation}
    \check{f}_{\mathcal{O}}(q,p) := \langle q,p | \hat{\mathcal{O}} | q,p \rangle \,.
\end{equation}
For a classical function $f(q,p)$, the affine quantization map \eqref{Af_formulae} give us a corresponding observer $A_{f}$. The dequantization map, then, returns another classical function $\check{f}(q,p)$,
\begin{equation}
\label{check_def}
    \check{f} (q,p) := \langle q,p | A_{f} | q,p \rangle \,.
\end{equation}
It corresponds to the average value of the function $f(q,p)$ with respect to the probability distribution of the phase space,
\begin{equation}
\label{probability_density}
    \rho_{\phi} (q,p) = \frac{1}{2\pi c_{\gamma}} |\langle q,p | \phi \rangle |^{2} \,
\end{equation}
in the case where $\phi = | q^{\prime}, p^{\prime} \rangle$. That is,
\begin{equation}
\label{check_f_rho_def}
    \check{f} (q,p) = \int_{0}^{\infty} \int_{-\infty}^{\infty} dq^{\prime} dp^{\prime} f(q^{\prime},p^{\prime}) \, \rho_{|q^{\prime}, p^{\prime} \rangle} (q,p) \,.
\end{equation}
We can obtain $\check{f}$ directly from the function $f$ using \eqref{Af_def} and \eqref{check_def}:
\begin{eqnarray}
\nonumber
     \check{f} (q,p) &=& \frac{1}{2\pi c_{\gamma}} \int_{0}^{\infty} \int_{-\infty}^{\infty} dq^{\prime}dp^{\prime} q^{\gamma} q^{\prime \,\gamma}  \int_{0}^{\infty} \int_{0}^{\infty} \frac{dx dx^{\prime}}{x^{\gamma +1} (x^{\prime})^{\gamma + 1}} f(q^{\prime},p^{\prime})
     \\
     \label{check_f}
     &\times& e^{ip(x-x^{\prime})} e^{-ip^{\prime}(x-x^{\prime})} \psi \left( \frac{x}{q} \right) \psi \left( \frac{x^{\prime}}{q} \right) \psi \left( \frac{x}{q^{\prime}} \right) \psi \left( \frac{x^{\prime}}{q^{\prime}} \right) \,.
\end{eqnarray}

This dequantization map, represented by a check mark above the function $\check{f} \,$, gives us a quantum phase-space (QPS) portrait, and it can be interpreted as a quantum correction of the classical function $f$. 
\begin{equation}
 \text{QPS} \,\, [{f}(q,p)] : f(q,p) \quad \mapsto \quad  A_{f} \quad \mapsto \check{f}(q,p) \,.
\end{equation}
It is worth mentioning that, when the constant $\hbar$ is reinstated, $\check{f} \rightarrow f$ for $\hbar \rightarrow 0$.

Comparing the quantum phase-space portrait to the Heisenberg approach, we notice that the QPS portrait of a classical function $f$ is not simply a measurement of the quantum observable $A_{f}$, which is the case for the expected value of the operator. While $\check{f}$ is calculated as a probablity distribution of the wavelets, the expected value of $A_{f}$ depends on the probability distribution of the energy eigenstates on the phase space,
\begin{equation}
    \langle \phi_{n} | A_{f} | \phi_{n}\rangle = \int_{0}^{\infty} \int_{-\infty}^{\infty} dq dp f(q,p) \, \rho_{\phi_{n}} (q,p) \,.
\end{equation}

With these tools in hand, let us apply this method for the case of black-hole gravity.

\section{Hamiltonian description of a black hole}
\label{Hamilt_descrip}

The classical theory of black holes is often studied with the Lagrangian formalism, considering the Einstein-Hilbert action from which Einstein's equations are derived. For our purposes, we will describe a spherical-symmetric black-hole solution with a Hamiltonian formalism, as proposed in \cite{Cavaglia.1996}.

Let us write the spherical symmetric solution of Einstein's equation as,
\begin{equation}
\label{BH metric}
    ds^{2} = -4 a(r) dt^{2} + 4 n(r) dr^{2} + b^{2}(r) d\Omega^{2} \,.
\end{equation}
Considering a $3+1$ decomposition of the space-time \cite{Gourgoulhon.2012}, the action is written as
\begin{equation}
    S = \frac{1}{16 \pi} \int_{V_{4}} d^{4}x \sqrt{-g}(R + 2\Lambda) - \frac{1}{8 \pi} \int_{\partial V_{4}} d^{3} x \sqrt{h} \boldsymbol{K} \,,
\end{equation}
where $\boldsymbol{K}$ is the extrinsic curvature of the spacial hypersurface, and $\Lambda$ is the cosmological constants. For simplicity, we are going to consider $\Lambda = 0$. With $g_{\mu\nu}$ given by \eqref{BH metric} and integrating in $\Omega$ \cite{Kuchar.1994}, the Lagrangian $\mathcal{L}$ of the action
\begin{equation}
    S = \int dt \int dr \mathcal{L} (a,b,l) \,,
\end{equation}
becomes
\begin{equation}
    \mathcal{L} = 2l \left( \frac{\dot{a}b \dot{b}}{l^{2}} + \frac{a \dot{b}^{2}}{l^{2}} + \frac{1}{4}\right) \,,
\end{equation}
with dots representing the derivative with respect to $r$. The non-dynamical quantity $l$ is given by $l = 4\sqrt{an}$. As usual, a simple calculation using the canonical momenta results in the Hamiltonian $\mathcal{H} = lH$, where
\begin{equation}
    H = \frac{1}{b^{2}} \left[p_{a} ( bp_{b} - ap_{a})\right] - \frac{1}{2} \,.
\end{equation}

Let us consider the canonical coordinate change,
\begin{eqnarray}
\label{alpha_beta}
\alpha = \ln(|a|) \quad ; \quad \beta = 2 \sqrt{|a|}b \,.
\end{eqnarray}
In these coordinates, the Hamiltonian is (Equation 2.25 of \cite{Cavaglia.1996}):
\begin{equation}
\label{Hamilt_Cavaglia}
    H = \frac{1}{2} \left[ \sigma \left( p_{{\beta}}^{2} - 4 \frac{p_{\alpha}^{2}}{\beta^{2}} \right) - 1 \right] \,,
\end{equation}
where $\sigma = \text{sign}(a)$. For a Schwarzschild black hole, $\sigma = 1$ means the region on the exterior of the black hole, since the metric has signature $(-, +, +, +)$, while $\sigma =-1$ represents the region on the interior of a black hole, where the coefficients for the time and radial coordinates change signs, that is, the signature becomes $(+, -, +, +)$. The coordinate $\beta$ spans from $0$ to $\infty$ in both regions, and the only way to differentiate the interior from the exterior is through $\sigma$.

We have $\alpha \in \mathbb{R}$ and $\beta \in \mathbb{R}_{+}$.  For the coordinate $\beta$, thus, the affine quantization is applicable. Meanwhile, the coordinate $\alpha$ seems an appealing candidate to play the role of time, imparting dynamics to the model. With this in mind, it is worth considering another canonical transformation for the coordinate $\alpha$:
\begin{equation}
\label{T,pT}
    T = \frac{\alpha}{p_{\alpha}} \quad; \quad P_{T} = \frac{p_{\alpha}^{2}}{2} \,.
\end{equation}
In these new coordinates, the Hamiltonian becomes,
\begin{equation}
    H = \frac{1}{2} \left[ \sigma \left( p_{{\beta}}^{2} - 8 \frac{p_{T}}{\beta^{2}} \right) - 1 \right] \,,
\end{equation}
The Hamiltonian constraint $H=0$ is, then, 
\begin{equation}
\label{Hamilt. const}
    H = \beta^{2} p_{{\beta}}^{2} - 8 p_{T}  - \sigma \beta^{2}  = 0 \,.
\end{equation}
This is the equation we will quantize in the next section.

If we had used the coordinates $(\alpha, \beta)$ instead of $(T, \beta)$, the quantization of the Hamiltonian constraint would lead to a Wheeler-DeWitt equation that, in principle, can be analytically solved. However, the lack of a dominant linear term in the differential equation could lead to a double arrow of time \cite{Bouhmadi.2020}. To avoid this, we described our model in a Schrödinger picture \cite{DeWitt.1967}, using a usual linearization strategy to obtain a linear term from a canonical transformation \cite{Vakili.2012}.

To further contextualize this set of coordinates, we can correlate $\beta$ to the volume of an Einstein-Rosen Bridge (ERB) connecting two black holes in a Thermo Field Dynamics (TFD) \cite{takahashi.1996, susskind.2016, tanaka.2022}. For late times, the volume $\mathcal{V}$ grows proportionally to the solid angle $\Omega^{2}$ \cite{nally.2019}. For the metric \eqref{BH metric}, the volume functional is given by (Equation (2.5) of \cite{nally.2019}):
\begin{equation}
    \mathcal{V}(b) = 2 \sqrt{|a(b)|} b^{3} \,.
\end{equation}
For small $b$, let us consider $|a| \rightarrow b^{-1}$. Then,
\begin{equation}
    \mathcal{V} \sim b^{2} \sim \beta^{4} \,.
\end{equation}
The coordinate $\beta$ is, therefore, related to the ERB volume, while $T$, as we hinted, will be identified as the entropic time parameter. Notice that if $\mathcal{V} = 0$ no bridge is formed, and we have a classical black hole with a singularity at its origin.

\section{Quantization of a black hole gravity}
\label{BH_quant}

Consider $\Psi(\beta, T)$ to be the black hole's wave-function. The quantization of the Hamiltonian constraint \eqref{Hamilt. const} leads to the Wheeler-DeWitt equation:
\begin{equation}
\label{WdW}
   \left[A_{\beta^{2} p_{{\beta}}^{2}} - 8 A_{p_{T}} - \sigma A_{\beta^{2}} \right] \Psi = 0 \,,
\end{equation}
where $A_{f}$ are as defined in Section \ref{AffQuant}. Before calculating those operators with the integral covariant method, we should analyze the Hilbert space of the wave function. Since equation \eqref{WdW} is separable, we must have $\Psi (\beta, T) = \Psi_{1} (\beta) \Psi_{2}(T)$, where
\begin{equation}
    \Psi_{1} \in \mathcal{H}_{-2}^{(\beta)} = L^{2} \left( \mathbb{R}_{+}, \beta d\beta\right) \quad; \quad \Psi_{2} \in \mathcal{H}_{-1}^{(T)} = L^{2} \left( \mathbb{R}, dT \right) \,.
\end{equation}
Thus, for the coordinate $T$, we can consider the canonical quantization given in \eqref{canonical_op.}. Then,
\begin{equation}
    A_{p_{T}}  = -i \partial_{T} \,.
\end{equation}
For the coordinate $\beta$, the volume coordinate, we will use the affine quantization. Let us apply the formalism laid out in Section \ref{AffQuant} with $\gamma =-2$. 

The operator $A_{\beta^{2}}$ is given in \eqref{quantiz_q}, 
\begin{equation}
    A_{\beta^{2}} =\frac{c_{0}}{c_{-2}} \beta^{2} \,.
\end{equation}
For $A_{\beta^{2} p_{{\beta}}^{2}}$, using equation \eqref{Af_formulae}, a lengthy calculation results in,
\begin{equation}
    A_{\beta^{2} p_{{\beta}}^{2}} = - \frac{c_{0}}{c_{-2}} \left[ \beta^{2}\partial_{\beta}^{2} + 3\beta \partial_{\beta} + \left(1 - \frac{c_{-2}^{(1)}}{ c_{0}} \right) \right] \,.
\end{equation}
Substituting these operators in the WdW equation \eqref{WdW}, we obtain a Schrödinger-like equation:
\begin{equation}
\label{Schrodinger}
    \frac{c_{0}}{8c_{-2}} \left[ \beta^{2}\partial_{\beta}^{2} + 3\beta \partial_{\beta} + \left( -\sigma \beta^{2} + 1 - \frac{c_{-2}^{(1)}}{c_{0}} \right) \right] \Psi = i \partial_{T} \Psi \,.
\end{equation}
Comparing \eqref{Schrodinger} to the usual Schrödinger equation, $\hat{H} \Psi = i (d/dt) \Psi$, we find the quantized Hamiltonian of the system,
\begin{equation}
\label{Hamilt_operator}
    \hat{H} = \frac{c_{0}}{8c_{-2}} \left[ \beta^{2}\partial_{\beta}^{2} + 3\beta \partial_{\beta} + \left( -\sigma \beta^{2} + 1 - \frac{c_{-2}^{(1)}}{c_{0}} \right) \right] \,.
\end{equation}
At his point, the canonical variable $T$ has no connection to the classical parameter $t$ of space-time. It is merely the phase of the wave function. However, following \cite{Caticha.2019}, we argue that the phase of the wave function equates to a Newtonian dynamics given by a universal time parameter. It is called entropic dynamics. With it, we recover the notion of an internal clock in both mechanical and informational senses. 

Another critical remark, the Hamiltonian operator depends on $\sigma$, that is, it changes for the black hole's interior and exterior regions, as expected.

The general solution of the WdW equation \eqref{Schrodinger} is,
\begin{equation}
\label{gen_solution}
    \Psi (\beta, T) = \beta^{-1} \left[ N_{1} J_{\nu} \left( \sqrt{-\sigma} \beta \right) + N_{2} Y_{\nu} \left( \sqrt{-\sigma} \beta \right) \right] e^{iET} \,,
\end{equation}
where $J_{\nu}$ and $Y_{\nu}$ are Bessel functions of the first and second kinds, respectively; $N_{1}, N_{2}$ are constants; $E$ is the energy of the system; and 
\begin{equation}
\label{nu}
    \nu = \left( \frac{c_{-2}^{(1)}}{c_{0}} + \frac{8c_{-2}}{c_{0}} E \right)^{\frac{1}{2}} \,.
\end{equation}
Before analyzing the boundary conditions, let us first specify $\sigma$.

\section{For $\sigma=-1$: the interior of the black hole}
\label{Interior_BH}

\subsection{Quantum solution for the interior of the BH}

The quantization of a gravitational model is useful to investigate domains in Planck-scale where gravity plays an essential role. In this paper, we will focus on the region around the singularity inside the event horizon. For this reason, we will choose $\sigma = -1$, which corresponds to the interior of the black hole.

Considering the solution \eqref{gen_solution} and the boundary conditions of the Hilbert space, we must have $N_{2} = 0$ and $\nu \in \mathbb{N}$, otherwise the solution is not well-defined at the origin ($\beta = 0$) \cite{Abramowitz.1964}. Therefore, renaming $\nu = n$, from \eqref{nu}, we obtain a discrete energy spectrum for the system:
\begin{equation}
\label{energy_level}
    E_{n} = \frac{c_{0}}{8c_{-2}} \left( n^{2} - \frac{c_{-2}^{(1)}}{c_{0}} \right) \,.
\end{equation}
And thus, the wave-function that describes the black-hole system is,
\begin{equation}
\label{general_wave_function}
    \Psi (\beta, T) = \sum_{n=0}^{\infty} \Psi_{n} = \sum_{n=0}^{\infty} N(n) \beta^{-1} J_{n} \left( \beta \right) e^{iE_{n}T} \,.
\end{equation}

To find the coefficients $N(n)$ of equation \eqref{general_wave_function}, let us calculate the norm of $\Psi_{n}$ for a  given $n$ \cite{gradshteyn.2007}.
\begin{equation}
    | \Psi_{n} |^{2} = |N(n)|^{2} \int_{0}^{\infty} \Psi_{n} \Psi^{*}_{n} \, \beta d\beta = |N(n)|^{2} \int_{0}^{\infty} \left[J_{\nu} (\beta ) \right]^{2} \frac{d\beta}{\beta} = \frac{|N(n)|^{2}}{2(n+1)} \,.
\end{equation}
As usual, we must have,
\begin{equation}
    \sum_{n=0}^{\infty} | \Psi_{n} |^{2} = \sum_{n=0}^{\infty} \frac{|N_{n}|^{2}}{2(n+1)} = 1 \,.
\end{equation}
We can choose, for example, $|N_{n}|^{2} = S_{p}^{-1} [2(n+1)^{-p+1}]$, where $p>1$, and $S_{p}$ is the convergent series defined as
\begin{equation}
    S_{p} := \sum_{k=1}^{\infty} \frac{1}{k^{p}} \,.
\end{equation}
And then,
\begin{equation}
    \Psi_{n} (\beta, T) = \sqrt{\frac{2}{S_p}} \frac{1}{(n+1)^{\frac{p+1}{2}}} \beta^{-1} J_{n} (\beta ) e^{iE_{n}T} \,.
\end{equation}

The energy levels can be calculated if we specify the fiducial vectors. Let us choose $\psi(\beta)$ as:
\begin{equation}
\label{fid_psi}
    \psi(\beta) = \frac{9}{\sqrt{6}} \beta^{\frac{3}{2}} e^{-\frac{3\beta}{2}} \,.
\end{equation}
Thus, $c_{-2} = c_{-1} = 1\,,c_{0} = 9/6$, and $c_{-2}^{(1)} = 9/8$. Substituting those values in equation \eqref{energy_level}, we obtain:
\begin{equation}
    E_{n} = \frac{3}{16} \left( n^{2} - \frac{3}{4} \right) \,.
\end{equation}
With this choice of fiducial vectors, the energy is negative for $n=0$ and positive for $n \geq 1$. This could indicate that the fundamental state $|\Psi_{0} \rangle$ is a wormhole solution, by direct relation to the violation of Null Energy Condition \cite{Visser.1995}.

\subsection{The Kantowski-Sachs metric}

According to Birkhoff's Theorem \cite{Birkhoff.1923}, every solution to the Einstein's equations of a spherical-symmetric gravitational field is the Schwarzschild solution (or Reissner-Nordström, if charged). Thus, the line element \eqref{BH metric} represents a Schwarzschild black hole, where
\begin{eqnarray}
\label{Schwarzchild_metric}
a(r) = \frac{1}{4} \left( 1- \frac{r_{S}}{r} \right) = n(r)^{-1} \quad ; \quad b(r) = r \,,
\end{eqnarray}
with $m$ being the black hole's mass and $r_S=2m$ its Schwarzschild radius. For the region inside the event horizon, that is, $r < 2m$, the coefficients $g_{tt}$ and $g_{rr}$ change their signs and the coordinate transformation $t \leftrightarrow r$ results in the homogenous and anisotropic Kantowski-Sachs cosmological solution \cite{Obregon.1998}:
\begin{align}
\label{Schw-KS}
    ds^2 = -\qty(\frac{2m}{t}-1)^{-1}dt^2 + \qty(\frac{2m}{t}-1)dr^2 + t^2 d\Omega^{2} \,.
\end{align}
This same structure can be found in metrics such as De Sitter and \cite{DeSitter.1917} and Taub-NUT \cite{Taub.1951,Newman.1963} spaces.

\begin{figure}[H]%
    \centering
    \includegraphics[scale=0.75]{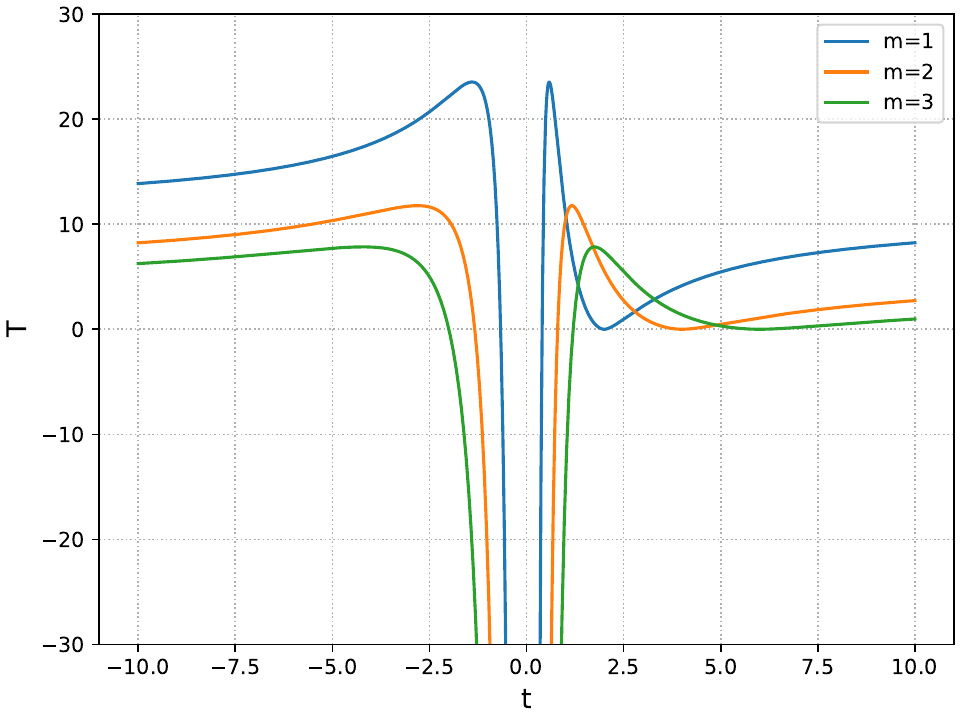}
    \caption{Entropic time $T$ versus cosmic time $t$ for different values of $m$.}%
    \label{Time_after_time}%
\end{figure}

With the Kantowski-Sachs metric, we can draw a connection between the entropic time $T$ and the cosmic time parameter $t$. The entropic time $T$ is given by equation \eqref{T,pT}, with $\alpha$ defined in equation \eqref{alpha_beta}, and $p_{\alpha}= -\dot{a}b^{2}/4$. \cite{Cavaglia.1996} Therefore,
\begin{equation}
\label{T=func(t)}
    T = \frac{\alpha}{p_{\alpha}} = -\frac{8}{m} \left( \frac{2m}{t} - 1 \right)^{2} \ln{\Big| \frac{m}{2t} - \frac{1}{4} \Big|} \,.
\end{equation}
Figure \ref{Time_after_time} shows the graphic of the entropic time as a function of the cosmic time. 

Notice that $T \rightarrow -\infty$ as $t \rightarrow 0$, but $T$ approaches a constant in late cosmic time. Indeed, the derivative of $T$ with respect to $t$ is,
\begin{equation}
\label{dT/dt=func(t)}
    \frac{dT}{dt} = \frac{32}{t^{2}} \left( \frac{2m}{t} - 1 \right)\ln{\Big| \frac{m}{2t} - \frac{1}{4} \Big|} - \frac{16}{(2m^{2}t-t^{2})} \left( \frac{2m}{t} - 1 \right)^{2} \,,
\end{equation}
In the limit $t\rightarrow \infty$, the derivative is null. This was to be expected. In the classical limit, the Hamiltonian constraint should hold, and the system should not have dynamics. Far from the origin, in the Kantwoski-Sachs metric represented by $t=0$, $T$ does not play a role in the system's evolution.

\section{Quantum phase-space portrait of a BH}
\label{QPS portrait}

With the quantum system described in the last section, we can employ the dequantization map defined in Subsection \ref{dequantization} to conduct a semi-classical analysis of this quantized model, decribed by the Hamiltonian operator \eqref{Hamilt_operator}. Classically, the identification of Hamiltonian as the total energy of the system gives us,
\begin{equation}
\check{H} = \langle \beta, p_{\beta} | \hat{H} | \beta, p_{\beta} \rangle = \mathcal{E} \,.
\end{equation}
Or, equivalently, 
\begin{equation}
\label{H=E}
   \check{\beta^{2} p_{{\beta}}^{2}} - \sigma \check{\beta^{2}} = 8E \,.
\end{equation}
Notice this equation is similar to the Hamiltonian constraint \eqref{Hamilt. const} if we call $p_{T} = E$. This identification comes from the entropic-dynamic understanding that $T$ is the clock parameter of the system. Equation \eqref{H=E} becomes, then, a quantum correction for the Hamiltonian constraint \eqref{Hamilt. const}.

Using \eqref{check_f}, we find that,
\begin{eqnarray}
\check{\beta^{2}} = \frac{c_{-3}c_{0}}{c_{-2}} q^{2} \, \quad;
\quad
\check{\beta^{2}p^{2}} &=& \frac{c_{-5}c_{0}}{c_{-2}} \left[ \beta^{2}p^{2} + \left( \frac{c_{-2}^{(1)} c_{-3}}{c_{-5}c_{0}} + \frac{c_{-5}^{(1)}}{c_{-5}} + \frac{c_{-3}}{c_{-5}} \right) \right] \,.
\end{eqnarray}
 Therefore, equation \eqref{H=E} becomes,
\begin{equation}
\label{check_Hamiltonian_constr.}
    \beta^{2}p_{\beta}^{2} - \sigma \beta^{2} = 8\frac{c_{-2}}{c_{0} c_{-5}} E - \left( \frac{c_{-2}^{(1)} c_{-3}}{c_{-5}c_{0}} + \frac{c_{-5}^{(1)}}{c_{-5}} + \frac{c_{-3}}{c_{-5}} \right) \,.
\end{equation}
This is the quantum phase-space portrait of the black-hole model; a classical equation obtained through the dequantization map provided by the affine quantization method.

Considering the fiducial vector given in \eqref{fid_psi}, we have $c_{3}=4/3$, $c_{-5}=40/9$, and $c_{-5}^{(1)}=3$. Therefore, 
\begin{equation}
\label{semi-clas_eq}
    \beta^{2}p_{\beta}^{2} - \sigma \beta^{2} = \frac{6}{5} E - \frac{47}{40} \,.
\end{equation}

We can get valuable insight from the phase-space portrait in a qualitative analysis of the model. For the interior region, that is, $\sigma=-1$, we have obtained a discrete energy spectrum in Section \ref{Interior_BH}, and thus there are different configurations for the phase-space of the system. As explained in Section \ref{Hamilt_descrip}, the coordinate $\beta$ is related to the volume of the ERB in this system. Then, if $\beta = 0$, no wormhole is formed, and the black hole has a singularity in its interior.

\begin{figure}[H]%
    \centering
    \includegraphics[scale=0.75]{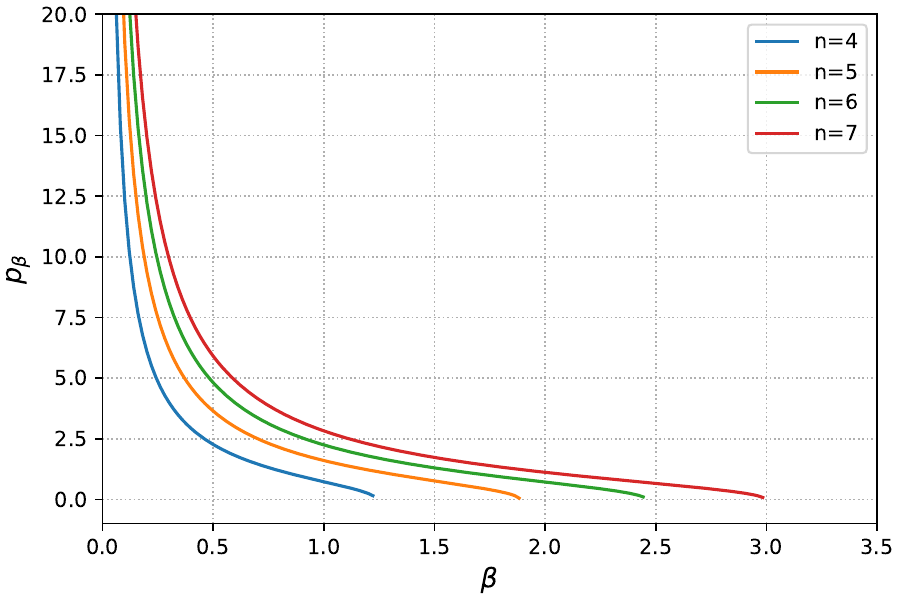}
    \caption{Phase-space portrait of the system for greater energy levels. The volume's contraction speeds near the origin, which remains an assymptote. For $\beta \rightarrow \infty$, the momentum tends to zero.}%
    \label{n>3}%
\end{figure}

Since we are concerned with the interior region, we have $\sigma =-1$. Then, for each energy level $n$, equation \eqref{semi-clas_eq} becomes,
\begin{equation}
\label{semi-clas_eq_En}
    \beta^{2}p_{\beta}^{2} + \beta^{2} = K_{n} \,.
\end{equation}
where $K_{n} = \frac{6}{5} E_{n} - \frac{47}{40}$. Let us start our analysis with the configurations for greater energy levels. For $n > 3$, $K_{n}>0$ and the solution behaves as the classical Schwarzschild model's phase-space \cite{Robertson.1968, Frolov.2011}, with eventual adjustments to the graphic's curvature (Figure \ref{n>3}). At least for higher energy levels, the singularity at the $r=0$ remains, and quantum effects do not seem relevant.

The situation changes for lower levels of energy. The lower energy states of a black hole system might represent the late stages of evaporation after the black hole has radiated most of its energy away. For $n=0,1,2,3$, we have $K_{n}<0$. Since we have a sum of square roots on the left side of equation \eqref{semi-clas_eq_En}, we must assume that $\beta$ is imaginary. Despite this, the volume $\mathcal{V} \sim \beta^{4}$ remains real and positive. Thus, it configures a physical situation.

We will interpret this situation later. For now, let us understand the behavior of the phase-space portrait for this case. For $\beta \in \mathbb{C}$, we have $p_{\beta} \in \mathbb{C}$. In fact, from the classical definition of the momenta,
\begin{equation}
\frac{dp_{\beta}}{dt} = - \frac{\partial H}{\partial \beta}  \,, \end{equation}
where $H$ is the Hamiltonian \eqref{Hamilt_Cavaglia}, we obtain
\begin{equation}
    p_{\beta} = -4\sigma \int \frac{p_{\alpha}}{\beta^{3}} dt \,.
\end{equation}
If we rename $\beta = i\lambda$, with $\lambda \in \mathbb{R}^{*}$, then,
\begin{equation}
    p_{\beta} = -4\sigma i \int \frac{p_{\alpha}}{\lambda^{3}} dt := i p_{\lambda} \,.
\end{equation}
The volume $\mathcal{V}$ can also be written in terms of $\lambda$: $\mathcal{V} \sim \beta^{4} = \lambda^{4}$. Therefore, without loss of generality, we can rewrite equation \eqref{semi-clas_eq_En} as
\begin{equation}
    \lambda^{2}p_{\lambda}^{2} - \lambda^{2} = K_{n} \,.
\end{equation}

\begin{figure}[H]%
    \centering
    \includegraphics[scale=0.75]{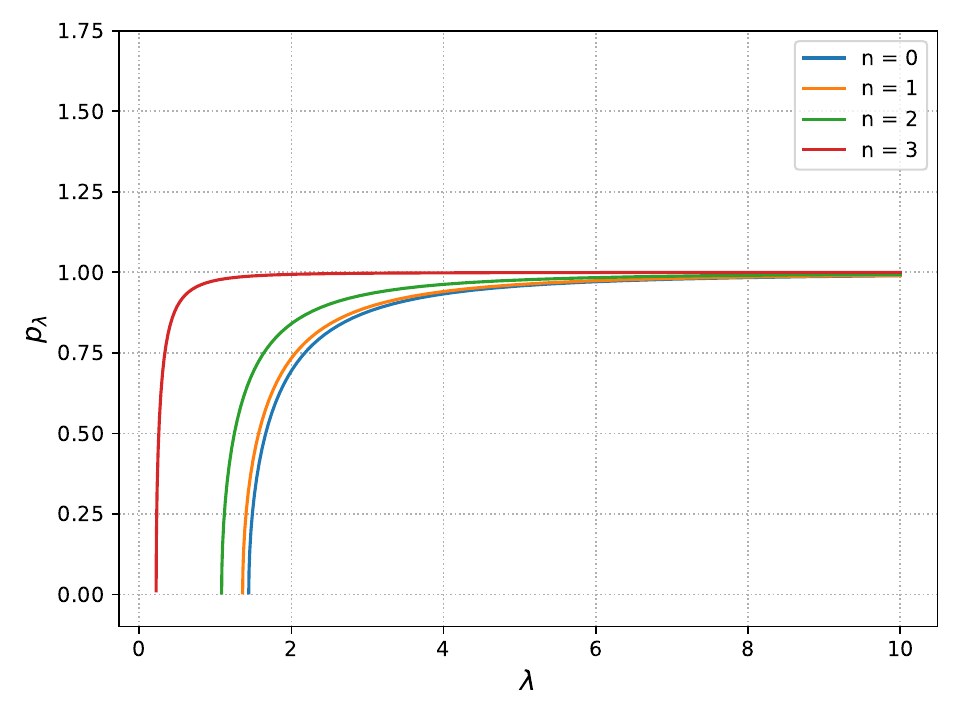}
    \caption{For lower levels of energy, we notice a formation of Einstein-Rosen bridges. The volume expands as the system loses energy.}%
    \label{n<4}%
\end{figure}

Figure \ref{n<4} shows the semi-classical phase space for energy levels $E_{n}$, $n=0,1,2,3$. The first four levels present a minimal, non-null value for $\lambda$, which means the formation of an Einstein-Rosen Bridge in the lower levels of energy. This result was foreshadowed in the quantum configuration, where we found that the ground level had negative energy; and also in the fact $\beta$ is imaginary, which indicates that there is not a semi-classical solution for the interior of the black hole, that is, there is not an event horizon.

In fact, considering the imaginary description of the space-time coordinates, that is, adopting an imaginary cosmic time, $t=i\tau$, an imaginary $\beta$ indicates a change of roles between the radial and time coordinates in the metric \eqref{BH metric}. To show this, let us recall the definition \eqref{alpha_beta}, $\beta = 2\sqrt{|a|}b$. For complex variables, $|a| = \sqrt{\Bar{a}a}$, where $\Bar{a}$ is the conjugate of $a$. Therefore, $\beta \in \mathbb{C} \Rightarrow b \in \mathbb{C}$. Thus, the coordinate change $(i\tau, b) \leftrightarrow (\tau, ib)$ holds. Classically, a change of roles in the time and radial coordinates happens when crossing the event horizon. This suggests that, for lower energy levels, \eqref{semi-clas_eq} has no solution to the interior of the black hole. That is, the system does not have an event horizon. It supports the result that a wormhole is formed in these cases.

We can see it analitically by observing how is the Hamiltonian \eqref{Hamilt_Cavaglia} for an imaginary $\beta$. If we rewrite the Hamiltonian \eqref{Hamilt_Cavaglia} in terms of $\lambda$, we obtain
\begin{equation}
    H = \frac{1}{2} \left[ \sigma \left( p_{{\beta}}^{2} - 4 \frac{p_{\alpha}^{2}}{\beta^{2}} \right) - 1 \right] = 
    \frac{1}{2} \left[ - \sigma \left( p_{{\lambda}}^{2} - 4 \frac{p_{\alpha}^{2}}{\lambda^{2}} \right) - 1 \right] \,.
\end{equation}
The Hamiltonian remains the same but with an inverted sign for $\sigma$, indicating that this is an outer region. Thus, the quantum-corrected Hamiltonian constraint \eqref{semi-clas_eq} does not have a solution for the interior of the black hole for lower energy states. That is, it does not have a horizon.

This semiclassical analysis is rich. For highly energetic black holes, for example, those formed by the collapse of supermassive stars, the behavior near the origin is as predicted by general relativity. However, if the black hole radiates more than consuming matter, it may evaporate and lose energy until it reaches the lower energy states. At this point, a wormhole is formed, and it expands as it loses mass. Because wormholes are known to be unstable \cite{gonzales.2008}, we assume they will eventually disappear, completing the evaporation cycle of the black hole.

It is important to remember that the energy levels are sensitive to the choice of fiducial vectors, both in the quantum and semi-classical portraits. A thorough analysis should reinstate the constants $G, \hbar$, and $c$ to properly interpret the energy levels.

\section*{Conclusion}

The quantization of a black hole's gravitational field can give us valuable insights into the enigmas surrounding those objects. In this initial investigation, we quantized a black hole model via the Wheeler-DeWitt equation in the mini super-space. We used the covariant integral quantization method, which we argued is more suitable for variables in different domains. Furthermore, it reproduces the canonical position and momentum operators for coordinates in $\mathbb{R}$. In particular, the quantization of strictly positive variables is given through the affine method, a covariant integral quantization that takes into consideration the affine symmetry of the system's phase space.

Our approach was to quantize the Hamiltonian constraint of the classical GR theory to obtain a Wheeler-DeWitt equation. We established that the evolution of the quantized system is given by an intrinsic variable that imparts dynamics to the model. This entropic time plays the role of a Newtonian clock, a universal time parameter. With the Kantowski-Sachs metric, we found the connection between the entropic time $T$ and the cosmic time $t$ inside the black hole. As expected, the entropic time action is dominant for early cosmic time but tends to a constant as the cosmic time grows. In other words, the universe's evolution according to the entropic time is null in the classical limit, and the Hamiltonian constraint holds. 

After identifying the time parameter, we recover a Schr{\"o}dinger-like equation, obtaining an effective Hamiltonian operator that governs the model. The solution of the Schrödinger-like Wheeler-DeWitt equation is a combination of polynomials, Bessel functions, and the time exponential. The system's energy spectrum is discrete in the black hole's interior. With our choice of fiducial vectors, the ground-state $|\Psi_{0} \rangle$ has negative energy, which suggests it could be a wormhole solution. 

The affine method offers an intuitive dequantization map in which we can recover a classical function of the position and momentum from a quantum operator. In the case of an operator obtained from a classical function, the quantum phase-space portrait can be interpreted as a quantum correction of said function. Through the dequantization map of the Hamiltonian operator, we presented a semi-classical analysis of the model. The discrete energy spectrum generates different phase-space configurations. For greater energy levels, the results are equivalent to what is expected from the classic general relativity theory. The indefinite contraction occurs, and the singularity remains at the origin. The situation changes for lower energy levels. These should correspond to the late stages of evaporation. In this situation, the solution has a minimum non-null volume and no event horizon, indicating that a wormhole has been formed.

The speculation left in this qualitative analysis of the semi-classical portrait will be further developed after we reinstate the fundamental constants $G, c$, and $\hbar$ to investigate the thermodynamical implications of the quantization and to probe how sensitive the energy is to the choice of fiducial vectors. Moreover, we will analyze the cases of micro black holes, a scenario in which quantum gravity becomes relevant near the horizon. A thermodynamical examination of those has recently been considered an alternative explanation for thermal-dominant gamma-ray bursts \cite{delBarco.2021}. Another goal would be to extend the results to the case of a charged black hole. The quantization of a black hole gravity opens many paths for future research.


\begin{thebibliography}{0}%
\makeatletter
\providecommand \@ifxundefined [1]{%
 \@ifx{#1\undefined}
}%
\providecommand \@ifnum [1]{%
 \ifnum #1\expandafter \@firstoftwo
 \else \expandafter \@secondoftwo
 \fi
}%
\providecommand \@ifx [1]{%
 \ifx #1\expandafter \@firstoftwo
 \else \expandafter \@secondoftwo
 \fi
}%
\providecommand \natexlab [1]{#1}%
\providecommand \enquote  [1]{``#1''}%
\providecommand \bibnamefont  [1]{#1}%
\providecommand \bibfnamefont [1]{#1}%
\providecommand \citenamefont [1]{#1}%
\providecommand \href@noop [0]{\@secondoftwo}%
\providecommand \href [0]{\begingroup \@sanitize@url \@href}%
\providecommand \@href[1]{\@@startlink{#1}\@@href}%
\providecommand \@@href[1]{\endgroup#1\@@endlink}%
\providecommand \@sanitize@url [0]{\catcode `\\12\catcode `\$12\catcode
  `\&12\catcode `\#12\catcode `\^12\catcode `\_12\catcode `\%12\relax}%
\providecommand \@@startlink[1]{}%
\providecommand \@@endlink[0]{}%
\providecommand \url  [0]{\begingroup\@sanitize@url \@url }%
\providecommand \@url [1]{\endgroup\@href {#1}{\urlprefix }}%
\providecommand \urlprefix  [0]{URL }%
\providecommand \Eprint [0]{\href }%
\providecommand \doibase [0]{https://doi.org/}%
\providecommand \selectlanguage [0]{\@gobble}%
\providecommand \bibinfo  [0]{\@secondoftwo}%
\providecommand \bibfield  [0]{\@secondoftwo}%
\providecommand \translation [1]{[#1]}%
\providecommand \BibitemOpen [0]{}%
\providecommand \bibitemStop [0]{}%
\providecommand \bibitemNoStop [0]{.\EOS\space}%
\providecommand \EOS [0]{\spacefactor3000\relax}%
\providecommand \BibitemShut  [1]{\csname bibitem#1\endcsname}%
\let\auto@bib@innerbib\@empty
\end{thebibliography}%


\begin{thebibliography}{99}
%
\bibitem[Hawking (1970)]{Hawking.1970}
Hawking, S; Penrose, R. The singularities of gravitational collapse and cosmology. {\em Proc. Roy. Soc. Lond. A} {\bf 1970}, {\bf 214}; pp. 529--548.
%
\bibitem[Hawking (1974)]{Hawking.1974}
Hawking, S. Black Hole explosion?. {\em Nature} {\bf 1974}, {\bf 248}; pp. 30--31.
%
\bibitem[Thiemann (1993)]{Thiemann.1993}
Thiemann, T.; Kastrup, H.A. Canonical quantization of spherically symmetric gravity
in Ashtekar’s self-dual representation. {\em Nuclear Physics B} {\bf 1933}, {\bf 399}; pp. 211--258.
%
\bibitem[Obregón (1998)]{Obregon.1998}
Obregón, O.; Ryan Jr., M.P. Quantum Planck Size Black Hole States Without a Horizon. {\em Modern Physics Letters A} {\bf 1998}, {\bf 13} {\em 40}; pp. 3251--3258.
%
\bibitem[Yajnik (1998)]{Yajnik.1998}
Yajnik, U.A.; Narayan, K. Canonical quantization inside the Schwarzschild black hole. {\em Classical and Quantum Gravity} {\bf1998}, {\bf 15}; pp. 1315--1321.
%
\bibitem[Bouhmadi-López (2020)]{Bouhmadi.2020}
Bouhmadi-López, M.; Brahma, S.; Chen, C.-Y.; Chen, P.; Yeom, D.-H. Annihilation-to-nothing: a quantum gravitational boundary condition for the Schwarzschild black hole. {\em Journal of Cosmology and Astroparticle Physics} {\bf 2020}, {\bf 11}; 002.
%
\bibitem[Casadio (2022)]{Casadio.2022}
Casadio, R. A quantum bound on the compactness. {\em The European Physical Journal C} {\bf2022}, {\bf 82} {\em 10}.
%
\bibitem[Campiglia (2007)]{Campiglia.2007}
Campiglia, M.; Gambini, R.; Pullin, J. Loop quantization of spherically symmetric
midi-superspaces. {\em Classical and Quantum Gravity} {\bf 2007}, {\bf 24}; pp. 3649--3672.
%
\bibitem[Modesto (2008)]{Modesto.2008}
Modesto, L. Black Hole Interior from Loop Quantum Gravity. {\em Advances in High Energy Physics} {\bf 2008}; 459290.
%
\bibitem[Gambini (2013)]{Gambini.2013}
Gambini, R.; Pullin, J. Loop quantization of the Schwarzschild Black Hole. {\em Physical Review Letters} {\bf 2013}, {\bf 110}; 211301.
%
\bibitem[Bojowald (2020)]{Bojowald.2020}
Bojowald, M. Black-Hole Models in Loop Quantum Gravity. {\em Universe} {\bf 2020}, {\bf 6} {\em 125}.
%
\bibitem[Twareque Ali (2005)]{Twareque_Ali.2005}
Twareque Ali, S; Englis, M. Quantization Methods: A guide to physicists and analysts. {\em Reviews in MAthematical Physics} {\bf 2005}, {\bf 17} {\em 4}; pp. 391--490.
%
\bibitem[Gazeau (2014)]{Gazeau.2014}
 Gazeau, J.-P.; Baldiotti, M.C.; Fresenda, R. Three examples of covariant integral quantization. In Third International Satellite Conference on Mathematical Methods in Physics, Proceedings of ICMP Conference, Londrina, Brazil, 21-26 October 2013; Proceedings of Science: 2014.
%
\bibitem[Aldrovandi (1990)]{Aldrovandi.1990}
Aldrovandi, R.; Galetti, D. On the structure of quantum phase space. {\em Journal of Mathematical Physics} {\bf 1990}, {\bf 31} {\em 12}; pp. 2987--2995.
%
\bibitem[Rundle (2021)]{Rundle.2021}
Rundle, R.P.; Everitt, M.J. Overview of the Phase Space Formulation of Quantum
Mechanics with Application to Quantum Technologies. {\em Advanced QUantum Technologies} {\bf 2021}, {\bf 4}; 2100016.
%
\bibitem[Bergeron (2014)]{Bergeron.2014}
Bergeron, H.; Dapor, A.; Gazeau, J.-P.; Malkiewicz, P. Smooth Big Bounce from Affine Quantization. {\em Physical Review D} {\bf 2014}, {\bf 89}, 083522.
%
\bibitem[Bergeron (2015)]{Bergeron.2015}
Bergeron, H.; Czuchry, E.; Gazeau, J.-P.; Malkiewicz, P.; Piechocki, W. Smooth quantum dynamics of the mixmaster universe. {\em Physical Review D} {\bf 2014}, {\bf 92}, 061302(R).
%
\bibitem[Frion (2019)]{Frion.2019}
Frion, E.; Almeida, C.R. Affine quantization of the Brans-Dicke theory: Smooth bouncing
and the equivalence between the Einstein and Jordan frames. {\em Physical Review D} {\bf 2014}, {\bf 99}, 023524.
%
\bibitem[Zonetti (2014)]{Zonetti.2014}
Zonetti, S. Affine quantization of black holes: Thermodynamics, singularity removal,
and displaced horizons. {\em Physical Review D} {\bf 2014}, {\bf 90}, 064046.
%
\bibitem[Gourgoulhon (2012)]{Gourgoulhon.2012}
Gourgoulhon, E. \textit{3+1 Formalism in General Relativity}, 1st. ed.; Springer: New York, U.S.A., 2012.
%
\bibitem[DeWitt (1967)]{DeWitt.1967}
DeWitt, B. Quantum Theory of Gravity. I. The Canonical Theory. {\em Physical Review} {\bf 1967}, {\bf 160} {\em 5}; pp. 1113--1148.
%
\bibitem[Isham (1993)]{Isham.1993}
Isham, C.J. Canonical Quantum Gravity and the Problem of Time. In: {\em Integrable Systems, Quantum Groups, and Quantum Field Theories} Ibort, L.A.; Rodríguez, M.A., Eds.; NATO ASI Series (Series C: Mathematical and Physical Sciences) {\bf 409}; Springer: Dordrecht, 1993; pp. 157--287.
%
\bibitem[Caticha (2011)]{Caticha.2011}
Caticha, A. Entropic Dynamics, Time, and Quantum Theory. {\em Journal of Physics A: Mathematical and Theoretical} {\bf 2011}, {\bf 44}; 225303.
%
\bibitem[Cavaglia (1996)]{Cavaglia.1996}
Cavaglià, M.; Alfaro, V.; Filippov, A.T. Hamiltonian Formalism for Black Holes and Quantization II.  {\em International Journal of Modern Physics D} {\bf 1996}, {\bf 5} {\em (3)}; pp. 227--250.
%
\bibitem[Kantowski(1963)]{Kantowski.1963}
Kantowski, R.; Sachs, R. K. Some Spatially Homogeneous Anisotropic Relativistic Cosmological Models, {\em Journal of Mathematical Physics} {\bf 1966}, {\bf 7} (3); pp 443--446.
%
\bibitem[Almeida (2018)]{Almeida.2018}
Almeida, C.R.; Bergeron, H.; Gazeau, J.-P.; Scardua, A.C. Three examples of quantum dynamics on the
half-line with smooth bouncing. {\em Annals of Physics} {\bf 2018}, {\em 392}; pp.206--228.
%
\bibitem[Gazeau (2009)]{Gazeau.2009}
Gazeau, J.-P. \textit{Coherent States in Quantum Physics}; Wiley-VCH: Weinheim, Germany, 2009.
%
\bibitem[Kucha\u{r}(1994)]{Kuchar.1994}
Kuchař, K. V. Geometrodynamics of Schwarzschild black holes. {\em Physical Review D} {\bf1994} {\bf50} (6); pp. 3961–-3981.
%
\bibitem[Vakili (2012)]{Vakili.2012}
Vakili, B. Scalar field quantum cosmology: A Schrödinger picture. {\em Physical Letters B} {\bf 2012}, {\bf718}; pp. 34--42.
%
\bibitem[Takahashi (1996)]{takahashi.1996} 
Takahashi, Y.; Umezawa, H. Thermo Field Dynamics. \textit{International Journal of Modern Physics B} {\bf 1996}, \textbf{10}(13n14); pp. 1755--1805.
%
\bibitem[Susskind (2016)]{susskind.2016} 
Susskind, L. Computational complexity and black hole horizons. \textit{Fortschr. Phys.} \textbf{2016}, \textbf{64}(1); pp. 24-–43.
%
\bibitem[Tanaka (2022)]{tanaka.2022}
Tanaka, I. Dissipation process in eternal black holes. \textit{J. Phys. Commun.} \textbf{2022}, \textbf{6}; 055015.
%
\bibitem[Nally (2019)]{nally.2019}
Nally, R. Stringy effects and the role of the singularity in
holographic complexity. \textit{JHEP} \textbf{2019}, \textbf{09}; 094.
%
\bibitem[Caticha (2019)]{Caticha.2019}
Caticha, A. The Entropic Dynamics Approach to Quantum Mechanics. {\em Entropy} {\bf 2019}, {\bf 21}; 943.
%
\bibitem[Abramowitz(1964)]{Abramowitz.1964}
Abramowitz, M.; Stegun, I.A. \textit{Handbook of Mathematical Functions with Formulas, Graphs, and Mathematical Tables}, 1st. ed.; U.S. Government Printing Office: Washington, D.C., U.S.A., 1964.
%
\bibitem[Gradshteyn (2007)]{gradshteyn.2007}
Gradshteyn, I.S.; Ryzhik, I.M. \textit{Table of Integrals, Series, and Products}. 7th ed.; Academic Press: California, United States, 2007.
%
\bibitem[Visser(1995)]{Visser.1995}
Visser, M. \textit{Lorentzian Wormholes: From Einstein to Hawking}; American Inst. of Physics: Maryland, U.S.A., 1995.
%
\bibitem[Birkhoff (1923)]{Birkhoff.1923}
Birkhoff, George David. \textit{Relativity and Modern Physics}; Harvard University Press: Cambridge, U.S.A., 1923. 
%
\bibitem[DeSitter (1917)]{DeSitter.1917}
De Sitter, W. On Einstein's Theory of Gravitation and its Astronomical Consequences. Third Paper., {\em Monthly Notices of the Royal Astronomical Society} {\bf 1917}, {\bf 78} (1);  pp 3–-28.
%
\bibitem[Taub (1951)]{Taub.1951}
 Taub, A. H. Annals of Mathematics {\bf 1951}, {\bf 53}; pp. 472--490.
%
\bibitem[Newman (1963)]{Newman.1963}
Newman, E.; Tamburino, L.; Unti, T. Empty‐Space Generalization of the Schwarzschild Metric, {\em Journal of Mathematical Physics} {\bf 1963}, {\bf 4} (7); pp 915--923.
%
\bibitem[Robertson (1968)]{Robertson.1968}
Robertson, H.P; Noonan, T.W. \textit{Relativity and Cosmology}; W. B. Saunders: Philadelphia, U.S.A.,  1968. 
%
\bibitem[Frolov (2011)]{Frolov.2011}
Frolov, V.P.; Zelnikov, A. {\em Introduction to Black Hole Physics}; Oxford University Press: Oxford, England, 2011.
%
%
%
\bibitem[Gonz{\'a}les (2008)]{gonzales.2008}
Gonz{\'a}lez, J.A.; Guzm{\'a}n, F.S.; Sarbach, O. Instability of wormholes supported by a ghost scalar field: I. Linear stability analysis. \textit{Classical and Quantum Gravity} \textbf{2008}, \textbf{26}: 015010.
%
\bibitem[del Barco (2021)]{delBarco.2021}
del Barco, O. Primordial black hole origin for thermal gamma-ray bursts. {\em Monthly Notices of the Royal Astronomical Society} {\bf 2021} {\bf 506}; pp. 806–-812.
%
%

\end{thebibliography}
\end{document}